\documentclass[showpacs,aps,prd,reprint,superscriptaddress,nofootinbib,longbibliography,preprintnumbers]{revtex4-1}
\usepackage[colorlinks=true, pdfstartview=FitV, linkcolor=magenta,citecolor=blue, urlcolor=magenta,
bookmarks=true, bookmarksnumbered=true, breaklinks]{hyperref}
\usepackage[dvipdfmx]{graphicx}
\usepackage{here}
\usepackage{url}
\usepackage{amsmath,amssymb,bm,color,longtable,mathrsfs,amsfonts,slashed}

\newcommand{\Slash}[1]{{\ooalign{\hfil/\hfil\crcr$#1$}}}

\begin{document}
\begin{flushright}
\end{flushright}

\preprint{RIKEN-iTHEMS-Report-22, YITP-22-127}

\title{Probing the hadron mass spectrum in dense two-color QCD with the linear sigma model}

\author{Daiki~Suenaga}
\email[]{daiki.suenaga@riken.jp}
\affiliation{Strangeness Nuclear Physics Laboratory, RIKEN Nishina Center, Wako 351-0198, Japan}
\affiliation{Research Center for Nuclear Physics,
Osaka University, Ibaraki 567-0048, Japan }

\author{Kotaro~Murakami}
\email[]{kotaro.murakami@yukawa.kyoto-u.ac.jp}
\affiliation{Yukawa Institute for Theoretical Physics, Kyoto University, Kyoto 606-8502, Japan}
\affiliation{Interdisciplinary Theoretical and Mathematical Sciences Program (iTHEMS), RIKEN, Wako 351-0198,
Japan}

\author{Etsuko~Itou}
\email[]{itou@yukawa.kyoto-u.ac.jp}
\affiliation{Research Center for Nuclear Physics,
Osaka University, Ibaraki 567-0048, Japan }
\affiliation{Interdisciplinary Theoretical and Mathematical Sciences Program (iTHEMS), RIKEN, Wako 351-0198,
Japan}
\affiliation{Department of Physics, and Research and Education Center for Natural Sciences, Keio University, 4-1-1
Hiyoshi, Yokohama, Kanagawa 223-8521, Japan}

\author{Kei~Iida}
\email[]{iida@kochi-u.ac.jp,}
\affiliation{Department of Mathematics and Physics, Kochi University, 2-5-1 Akebono-cho, Kochi 780-8520, Japan}

\date{\today}

\begin{abstract}

We investigate modifications of hadron masses at finite quark chemical potential in two-flavor and two-color QCD, of which data are available from lattice simulations, within a linear sigma model based on approximate Pauli-Gursey $SU(4)$ symmetry. The model describes not only ground-state scalar diquarks and pseudo-scalar mesons but also the excited pseudo-scalar diquarks and scalar mesons; each ground-state diquark (meson) has the corresponding excited diquark (hadron) with opposite parity as a chiral partner. Effects of chiral symmetry breaking and diquark condensates are incorporated by a mean-field treatment. We show that various mixings among the hadrons, which are triggered by the breakdown of baryon number conservation in the superfluid phase, lead to a rich hadron mass spectrum. We discuss the influence of $U(1)_A$ anomaly on the density dependence of the mass spectrum and also manifestations of the chiral partner structures as the density increases in the superfluid phase. The predicted hadron masses are expected to provide future lattice simulations with useful information on such symmetry properties in dense two-color QCD.
 \end{abstract}


\maketitle

\section{Introduction}
\label{sec:Introduction}

Toward understanding Quantum Chromodynamics (QCD) at finite quark chemical potential $\mu_q$, two-color QCD (QC$_2$D) with even number of quark flavors is useful since in such a QCD-like theory {\it lattice QCD} simulations work even at finite $\mu_q$ without suffering from the so-called {\it sign problem}~\cite{Muroya:2003qs,Aarts:2015tyj}. Based on this advantage, so far many efforts from lattice QCD simulations at finite $\mu_q$ in QC$_2$D have been devoted to understanding of, e.g., modifications of hadron masses, gluon propagators, phase diagram of QC$_2$D, electromagnetic transport coefficients, and so on~\cite{Hands:1999md,Kogut:2001na,Hands:2001ee,Muroya:2002ry,Chandrasekharan:2006tz,Hands:2006ve,Hands:2007uc,Hands:2010gd,Cotter:2012mb,Hands:2012yy,Boz:2013rca,Braguta:2016cpw,Puhr:2016kzp,Boz:2018crd,Astrakhantsev:2018uzd,Iida:2019rah,Wilhelm:2019fvp,Buividovich:2020gnl,Iida:2020emi,Astrakhantsev:2020tdl,Bornyakov:2020kyz,Buividovich:2020dks,Buividovich:2021fsa,Iida:2022hyy}. Therefore, lattice simulations in QC$_2$D at finite $\mu_q$ serve as a {\it numerical experiment} for future investigation of dense QCD.

Although lattice simulations are powerful, they only provide us with numerical information. In this regard, examinations of the simulation results based on effective models give us deeper insights into dense QCD. Motivated by this fact, hadron mass modifications and phase structures at finite $\mu_q$ were theoretically investigated within chiral perturbation theory~\cite{Kogut:1999iv,Kogut:2000ek,Lenaghan:2001sd,Kanazawa:2009ks,Adhikari:2018kzh}, hidden local symmetry (HLS)~\cite{Harada:2010vy}, Nambu-Jona-Lasinio (NJL) type model~\cite{Ratti:2004ra,Sun:2007fc,Fukushima:2007bj,Brauner:2009gu,Andersen:2010vu,Zhang:2010kn,He:2010nb,Imai:2012hr,Duarte:2015ppa,Chao:2018czo,Khunjua:2020xws,Khunjua:2021oxf}, and quark-meson coupling model with the functional method~\cite{Strodthoff:2011tz,Strodthoff:2013cua,Khan:2015puu}. Delineation of gluon propagators and transport coefficients in dense QC$_2$D was also attempted by using the Dyson-Schwinger equation~\cite{Contant:2019lwf} as well as by combining a massive gluon model with quasiparticle description of quarks ~\cite{Suenaga:2019jjv,Kojo:2021knn,Suenaga:2021bjz}. In addition to those field-theoretical approaches, which are broadly employed, recently a unified picture that connects the smooth transition from hadronic matter to quark matter with the quark model has been developed~\cite{Kojo:2021hqh,Kojo:2022psi}. 
From these studies, it is expected that a deeper understanding of dense QC$_2$D properties, most of which are commonly shared by three-color QCD, is achieved~\cite{Baym:2017whm}.

In QC$_2$D, diquarks made of two quarks form a color singlet and hence can be regarded as baryons. Accordingly, baryonic matter is formed as a many-body system of diquarks obeying the Bose-Einstein statistics. As a result, when the baryon chemical potential or, equivalently, the quark chemical potential $\mu_q$ exceeds a certain critical value, the Bose-Einstein condensate (BEC) phase of diquarks emerges at sufficiently low temperature~\cite{Kogut:1999iv,Kogut:2000ek}. The phase is often called the {\it diquark condensed phase} or the {\it baryon superfluid phase} since the baryon number conservation is violated here. In contrast, the normal phase with no BECs, which is continuously connected to the vacuum, i.e., the system with vanishing temperature and chemical potential, is simply referred to as the {\it hadronic phase}. In this phase, all thermodynamic quantities are independent of $\mu_q$, which is known as the {\it Silver Blaze property}.

Emergence of the baryon superfluidity is manifestly reflected by hadron mass spectrum. For instance, the baryon number violation in the superfluid phase causes mixings among mesons and diquarks having identical quantum numbers~\cite{Ratti:2004ra}. The appearance of the Nambu-Goldstone (NG) bosons in association with the breakdown of $U(1)_B$ baryon number symmetry is another striking consequence~\cite{Kogut:1999iv,Kogut:2000ek}. For this reason, lattice simulations that reveal modifications of hadron masses in the baryon superfuid phase as well as in the hadronic phase were performed by several groups~\cite{Hands:1999md,Kogut:2001na,Muroya:2002ry,Hands:2007uc,Wilhelm:2019fvp,Murakami2022}. In particular, in Ref.~\cite{Murakami2022} the simulation was extended in such a way as to include not only the ground-state hadrons but also the orbitally excited ones having opposite parities.

Motivated by the above progress in lattice studies, in the present study, we theoretically investigate hadron mass modifications in both the hadronic and baryon superfluid phases at zero temperature by utilizing a linear sigma model~\cite{Gell-Mann:1960mvl}. Since the linear sigma model is based on the linear representation of quarks, the model has two noteworthy advantages from the symmetry point of view: (i) The model can describe both the ground-state hadrons and excited ones in a unified way, which allows us to identify the {\it chiral partners}. (ii) The model can incorporate changes of the ground-state configurations associated with in-medium chiral symmetry restoration in a broad range of $\mu_q$  at mean-field level~\cite{Hatsuda:1994pi}.\footnote{Investigation of modifications of light hadron masses from the aspect of chiral restoration at finite density through the linear sigma model of three-color QCD have been done widely by several methods~\cite{Schaefer:2006ds,Tiwari:2012yy,Herbst:2013ail,Sakai:2013nba,Tawfik:2014gga,Tawfik:2015tga,Fejos:2017kpq,Fejos:2018dyy,Suenaga:2019urn,Tawfik:2019rdd} such as functional methods. The model has also been applied to QCD with an isospin chemical potential~\cite{Mao:2006zr}.} In particular, we concentrate on spin zero hadrons in this exploratory work where inputs are provided by the recent lattice results~\cite{Murakami2022}. Then, we demonstrate how symmetry properties related to chiral symmetry and $U(1)_A$ axial anomaly in dense QC$_2$D are extracted by the mass spectrum. Moreover, we present predictions of novel hadron mass modifications, which might provide useful information on the symmetry insights of dense QC$_2$D for future lattice simulations.

This article is organized as follows. In Sec.~\ref{sec:Model}, general properties of QC$_2$D are briefly explained, and accordingly the linear sigma model to investigate hadron mass modifications at finite $\mu_q$ is introduced. In Sec.~\ref{sec:Input}, input information from recent lattice simulations is presented, and we therefrom examine the $\mu_q$ dependence of mean fields to delineate the emergence of the baryon superfluid phase. In Sec.~\ref{sec:MassSpectrum} the resultant hadron mass spectra at finite $\mu_q$ are demonstrated by focusing on effects of $U(1)_A$ anomaly, and also discussions on the chiral partner structures are provided. In Sec.~\ref{sec:Conclusions} we conclude the present work.

\section{Model}
\label{sec:Model}

In this section, we construct our linear sigma model from symmetry arguments.

\subsection{General properties of QC$_2$D}
\label{sec:General}

Within the linear sigma model, hadron states are provided by the linear representation of quark bilinears sharing the same symmetry properties, which in turn determine the structure of hadron interactions. In QC$_2$D with two flavors ($N_f=2$), the flavor symmetry is characterized by the Pauli-Gursey $SU(4)$ symmetry~\cite{Smilga:1994tb,Splittorff:2000mm,Kogut:1999iv,Kogut:2000ek} rather than $SU(2)_L\times SU(2)_R\times U(1)_B$. In this subsection, before presenting our linear sigma model we briefly review emergence of the Pauli-Gursey $SU(4)$ symmetry by turning back to the fundamental QC$_2$D Lagrangian.

The QC$_2$D Lagrangian for massless two quarks ($N_f=2$) is of the form
\begin{eqnarray}
{\cal L}_{\rm QC_2D} = \bar{\psi}i\Slash{D}\psi \ , \label{QC2DLagrangian}
\end{eqnarray}
where $\psi=(u,d)^T$ is the quark doublet and $D_\mu\psi = \partial_\mu\psi+ig_c A_\mu^aT_c^a\psi$ is the covariant derivative describing interactions between the quarks and gluons, with $T_c^a=\tau_c^a/2$ being the $SU(2)_c$ generator ($\tau_c^a$ is the Pauli matrix in color space). Introducing the Weyl representation for the Dirac matrices for convenience, one can express the Lagrangian~(\ref{QC2DLagrangian}) in terms of left-handed and right-handed quarks as
\begin{eqnarray}
{\cal L}_{\rm QC_2D} &=& \psi^\dagger_R i\partial_\mu\sigma^\mu\psi_R -g_c\psi_R^\dagger A_\mu^aT_c^a\sigma^\mu \psi_R \nonumber\\
&+& \psi^\dagger_L i\partial_\mu\bar{\sigma}^\mu\psi_L -g_c\psi_L^\dagger A_\mu^aT_c^a\bar{\sigma}^\mu \psi_L \ . \label{QC2DLagrangian2}
\end{eqnarray}
In this Lagrangian, we have used $u=(u_R,u_L)^T$ and $d=(d_R,d_L)^T$ in the Weyl representation, and defined two component matrices $\sigma^\mu=({\bm 1}, \sigma^i)$ and $\bar{\sigma}^\mu=({\bm 1},-\sigma^i)$ in spinor space with the Pauli matrix $\sigma^i$. Here, we utilize the pseudo-reality of the Pauli matrix to obtain the relations
\begin{eqnarray}
T_c^a = -\tau_c^2(T_c^a)^T\tau_c^2\ , \ \ \sigma^i = -\sigma^2(\sigma^i)^T\sigma^2\ ,
\end{eqnarray}
and accordingly introduce ``conjugate quarks'' 
\begin{eqnarray}
\tilde{\psi}_R \equiv \sigma^2\tau_c^2\psi_R^* \ , \ \ \tilde{\psi}_L \equiv \sigma^2\tau_c^2\psi_L^*\ .
\end{eqnarray}
Then, the Lagrangian~(\ref{QC2DLagrangian2}) can be expressed in a unified form as
\begin{eqnarray}
{\cal L}_{\rm QC_2D} = \Psi^\dagger i\partial_\mu\sigma^\mu\Psi-g_c\Psi^\dagger A_\mu^aT_c^a\sigma^\mu\Psi\ , \label{QC2DFour} 
\end{eqnarray}
when we introduce the four-component column vector as
\begin{eqnarray}
\Psi \equiv \left(
\begin{array}{c}
\psi_R \\
\tilde{\psi}_L \\
\end{array}
\right) =  \left(
\begin{array}{c}
u_R \\
d_R \\
\tilde{u}_L \\
\tilde{d}_L \\
\end{array}
\right) \ . \label{PsiFour}
\end{eqnarray}

The Lagrangian~(\ref{QC2DFour}) is obviously invariant under $SU(4)$ transformation of $\Psi$:
\begin{eqnarray}
\Psi \to g \Psi\ , \label{PsiTrans}
\end{eqnarray}
with $g\in SU(4)$,\footnote{We consider only $SU(4)$ symmetry but not $U(4)$ one, since the $U(1)_A$ axial symmetry is explicitly broken due to the anomaly.} rather than $SU(2)_L\times SU(2)_R\times U(1)_B$ chiral transformation. This extended symmetry is sometimes referred to as the Pauli-Gursey symmetry. As can be seen from Eq.~(\ref{PsiFour}), the Pauli-Gursey symmetry is realized by treating $\psi$ and the ``conjugate quarks'' $\tilde{\psi}$ as one quartet. 

The baryon number symmetry, i.e., quark number symmetry is embedded in the $SU(4)$ symmetry. In fact, from Eq.~(\ref{PsiFour}) the quark number transformation reads
\begin{eqnarray}
\Psi \to {\rm e}^{-i\theta_q { J}} \Psi\  \ \ {\rm with} \ \  \ { J} \equiv \left(
\begin{array}{cc}
{\bm 1} & 0 \\
0 & -{\bm 1} \\
\end{array}
\right) \ . \label{BaryonPsi}
\end{eqnarray}
On the other hand, the $U(1)_A$ transformation is generated by a unit matrix as
\begin{eqnarray}
\Psi \to {\rm e}^{-i\theta_A { I}} \Psi\  \ \ {\rm with} \ \  \ { I} \equiv \left(
\begin{array}{cc}
{\bm 1} & 0 \\
0 & {\bm 1} \\
\end{array}
\right) \ . \label{AxialPsi}
\end{eqnarray}

One of the most characteristic properties of QC$_2$D is the symmetry breaking pattern triggered by a chiral condensate $\langle\bar{\psi}\psi\rangle$. Using Eq.~(\ref{PsiFour}) one can check 
\begin{eqnarray}
\bar{\psi}\psi   =-\frac{1}{2}(\Psi^T\sigma^2\tau_c^2E\Psi  + \Psi^\dagger\sigma^2\tau_c^2E^T\Psi^*) \ , \label{ChiralCond}
\end{eqnarray}
with
\begin{eqnarray}
 { E} \equiv  \left(
\begin{array}{cc}
0 & {\bm 1} \\
-{\bm 1} & 0 \\
\end{array}
\right) \,
\end{eqnarray}
the $4\times4$ symplectic matrix. Equation~(\ref{ChiralCond}) implies that the chiral condensate $\langle\bar{\psi}\psi\rangle$ is invariant under transformations satisfying 
\begin{eqnarray}
h^T { E}h = { E}\ , \label{SympDef}
\end{eqnarray}
where $h$ is an element of $Sp(4)$ belonging to a subgroup of the original $SU(4)$. In other words, the symmetry breaking pattern triggered by the chiral condensate is $SU(4)\to Sp(4)$~\cite{Kogut:1999iv,Kogut:2000ek}. 

Based on the general properties of QC$_2$D presented in this subsection, we introduce hadron fields from the quark bilinears in Sec.~\ref{sec:Bilinear} and construct the linear sigma model in such a way as to respect $SU(4)$ symmetry in Sec.~\ref{sec:LSM}.

\subsection{The flavor matrix $\Sigma$}
\label{sec:Bilinear}

The advantage of employing the linear sigma model of QCD is that we can simultaneously investigate the ground-state hadrons and their chiral partners having opposite parities, i.e., the $P$-wave excited states in the quark-model sense, on an equal footing~\cite{Gell-Mann:1960mvl}. Such an advantage is explicitly implemented by introducing a flavor matrix $\Sigma$ containing the hadrons defined through a quark bilinear field in the linear representation. Here, we introduce the corresponding $4\times4$ $\Sigma$ matrix in QC$_2$D and present its properties.

As shown in Sec.~\ref{sec:General}, QC$_2$D has the Pauli-Gursey $SU(4)$ symmetry when the quarks are massless, and accordingly any hadronic theory has to respect the symmetry as well. Hence, one useful definition of $\Sigma$ in terms of the quark bilinear field may be\footnote{Here, the symbol ``$\sim$" denotes the correspondence between the composite state in the linear sigma model and the quark bilinear in QC$_2$D. }
\begin{eqnarray}
\Sigma_{ij} \sim \Psi^T_j\sigma^2\tau_c^2\Psi_i  \ . \label{SigmaJ}
\end{eqnarray}
This $\Sigma$ is a flavor $4\times4$ matrix labeled by $i$ and $j$, where the summations over spinor and color indices are implicitly done. The flavor matrix $\Sigma$ is antisymmetric as $\Sigma = -\Sigma^T$ due to the Grassman nature of $\Psi$. One can see that $\Sigma$ transforms homogeneously under the $SU(4)$ as
\begin{eqnarray}
\Sigma \to g\Sigma g^T\ , \label{SigmaTrans}
\end{eqnarray}
from the linear transformation law of the quartet in Eqs.~(\ref{PsiFour}) and~(\ref{PsiTrans}). More explicitly, the $\Sigma$ in Eq.~(\ref{SigmaJ}) takes the form of
\begin{eqnarray}
&& \Sigma \sim\nonumber\\
 &&  \left(
\begin{array}{cccc}
0 &d_R^T\sigma^2\tau_c^2  u_R & u_L^\dagger u_R & d_L^\dagger u_R \\
-d_R^T \sigma^2\tau_c^2 u_R &0 & {u}_L^\dagger d_R & {d}_L^\dagger d_R \\
- u_L^\dagger u_R & -u_L^\dagger d_R & 0 & {d}_L^\dagger \sigma^2\tau_c^2{u}^*_L \\
- {d}_L^\dagger u_R & -{d}_L^\dagger d_R& - {d}_L^\dagger\sigma^2\tau_c^2 {u}^*_L & 0\\
\end{array}
\right) . \nonumber\\
\label{SigmaMatrixIJ}
\end{eqnarray}
Equation~(\ref{SigmaMatrixIJ}) implies that mesons and diquark baryons can be treated in a unified way. To see such a structure more clearly, we try to rewrite the matrix~(\ref{SigmaMatrixIJ}) in terms of hadronic states. For this reason, we define interpolating fields for the hadrons by
\begin{eqnarray} 
\sigma &\sim& \bar{\psi}\psi = u_L^\dagger u_R + d_L^\dagger d_R + u^\dagger_R u_L+d_R^\dagger d_L\ , \label{SigmaInt}
\end{eqnarray}
\begin{eqnarray}
a_0^+ &\sim&\frac{1}{\sqrt{2}} \bar{\psi}\tau_f^-\psi = \sqrt{2} (d_L^\dagger u_R+d_R^\dagger u_L) \ ,\nonumber\\
a_0^{\bm -} &\sim& \frac{1}{\sqrt{2}}\bar{\psi}\tau_f^+\psi = \sqrt{2} (u_L^\dagger d_R+u_R^\dagger d_L)\ , \nonumber\\
a_0^0 &\sim& \bar{\psi}\tau_f^3\psi =  u_L^\dagger u_R-d_L^\dagger d_R+u^\dagger_R u_L-d_R^\dagger d_L\ , 
\end{eqnarray}
\begin{eqnarray}
\eta &\sim& \bar{\psi}i\gamma_5\psi =  i(u_L^\dagger u_R+d_L^\dagger d_R-u^\dagger_R u_L-d_R^\dagger d_L)\, ,
\end{eqnarray}
\begin{eqnarray}
\pi^+ &\sim&\frac{1}{\sqrt{2}} \bar{\psi}i\gamma_5\tau_f^-\psi = \sqrt{2} i(d_L^\dagger u_R-d_R^\dagger u_L) \ ,\nonumber\\
\pi^{\bm -} &\sim& \frac{1}{\sqrt{2}}\bar{\psi}i\gamma_5\tau_f^+\psi = \sqrt{2} i(u_L^\dagger d_R-u_R^\dagger d_L)\ , \nonumber\\
\pi^0 &\sim& \bar{\psi}i\gamma_5\tau_f^3\psi =  i(u_L^\dagger u_R-d_L^\dagger d_R-u^\dagger_R u_L+d_R^\dagger d_L)\ , \nonumber\\ \label{PiInt}
\end{eqnarray}
\begin{eqnarray}
B &\sim& -\frac{i}{\sqrt{2}} \psi^TC\gamma_5\tau_c^2\tau_f^2\psi \nonumber\\
&=& -\sqrt{2}i(d_R^T\sigma^2\tau^2u_R + d_L^T\sigma^2\tau^2u_L) \, ,
  \nonumber\\ 
\bar{B} &\sim& -\frac{i}{\sqrt{2}}\psi^\dagger C\gamma_5\tau_c^2\tau_f^2\psi^* \nonumber\\
&=& - \sqrt{2}i(d_R^\dagger\sigma^2\tau^2u_R^* + d_L^\dagger\sigma^2\tau^2u_L^*)   \, , \label{BInt} 
\end{eqnarray}
\begin{eqnarray}
B' &\sim&- \frac{1}{\sqrt{2}} \psi^TC\tau_c^2\tau_f^2\psi \nonumber\\
&=& - \sqrt{2}(d_R^T\sigma^2\tau^2u_R - d_L^T\sigma^2\tau^2u_L) \ ,
  \nonumber\\ 
\bar{B}' &\sim& \frac{1}{\sqrt{2}}\psi^\dagger C\tau_c^2\tau_f^2\psi^* \nonumber\\
&=&   \sqrt{2}(d_R^\dagger\sigma^2\tau^2u_R^* - d_L^\dagger\sigma^2\tau^2u_L^*)  \ , \label{BpInt}
\end{eqnarray}
with $\tau_f^\pm = \tau_f^1\pm i\tau_f^2$ ($\tau_f^a$ is the Pauli matrix in flavor space) and $C=i\gamma^2\gamma^0$ the charge-conjugation operator. For the mesons defined in Eqs.~(\ref{SigmaInt})~-~(\ref{PiInt}), we have employed the notations which are ordinarily adopted in three-color QCD, and thus their chiral properties are well known. For the baryons defined in Eqs.~(\ref{BInt}) and~(\ref{BpInt}), current structures are largely different from the case of three-color QCD where baryons are composed of three quarks; $B$ and $\bar{B}$ represent diquark and antidiquark baryons, respectively, which are singlet in both spin and isospin and characterized by $J^P=0^+$, while $B'$ and $\bar{B}'$ are their chiral partners carrying opposite parities. In fact, $B$ ($\bar{B}$) and $B'$ ($\bar{B}'$) are interchanged under the $SU(2)_A$ axial transformation. In order to manifestly display the properties, we tabulate quantum numbers of the hadrons in Table~\ref{tab:Assignment}.
\begin{table}[htbp]
\begin{center}
  \begin{tabular}{c|ccc} \hline
Hadron & $J^P$ & Quark number & Isospin \\ \hline 
$\sigma$ & $0^+$ & $0$ & $0$ \\
$a_0$ & $0^+$ & $0$ & $1$ \\
$\eta$ & $0^-$ & $0$ & $0$ \\
$\pi$ & $0^-$ & $0$ & $1$ \\
$B$ ($\bar{B}$) & $0^+$ & $+2 (-2)$ & 0 \\
$B'$ ($\bar{B}'$) & $0^-$ & $+2 (-2)$ & 0  \\ \hline
 \end{tabular}
\caption{The quantum numbers carried by the hadrons defined in Eqs.~(\ref{SigmaInt})~-~(\ref{BpInt}). }
\label{tab:Assignment}
\end{center}
\end{table}

Using the hadronic states defined in Eqs.~(\ref{SigmaInt})~-~(\ref{BpInt}), the matrix~(\ref{SigmaMatrixIJ}) can be described in terms of the hadrons as
\begin{widetext}
\begin{eqnarray}
\Sigma =   {\cal N}\left(
\begin{array}{cccc}
0 & -B'+iB &\frac{\sigma-i\eta+a^0-i\pi^0}{\sqrt{2}} & a^+-i\pi^+ \\
 B'-iB & 0 & a^--i\pi^- & \frac{\sigma-i\eta-a^0+i\pi^0}{\sqrt{2}} \\
-\frac{\sigma-i\eta+a^0-i\pi^0}{\sqrt{2}} & -a^-+i\pi^- & 0 & - \bar{B}' + i\bar{B} \\
-a^++i\pi^+&-\frac{\sigma-i\eta-a^0+i\pi^0}{\sqrt{2}} & \bar{B}'-i\bar{B}& 0 \\
\end{array}
\right)  \ . \label{SigmaPAssign}
\end{eqnarray}
\end{widetext}
As for the normalization constant ${\cal N}$, we take ${\cal N}=1/2$ for later use.

 The matrix~(\ref{SigmaPAssign}) implies that, when $\sigma$ is replaced by its mean field $\sigma_0$ responsible for the chiral condensate $\langle \bar{\psi}\psi\rangle$, the vacuum expectation value (VEV) of $\Sigma$ is proportional to the symplectic matrix
\begin{eqnarray}
\langle\Sigma\rangle_{\rm ch.} \, {\propto}\,  { E}\ . \label{PhiVEV}
\end{eqnarray}
Thus, from the transformation law in Eq.~(\ref{SigmaTrans}) we see that the VEV $\langle\Sigma\rangle_{\rm ch.}$ is singlet only when $g$ is replaced by $h$ satisfying Eq.~(\ref{SympDef}), which obviously reflects the symmetry breaking pattern of $SU(4)\to Sp(4)$ as explained in the end of Sec.~\ref{sec:General}.

The matrix~(\ref{SigmaPAssign}) can be written in a more compact form. In fact, once one defines $S^a$, $P^a$, $B^i$, and $B'^i$ as
\begin{eqnarray}
&& \sigma = S^0\ , \ \ a_0^0 = S^3\ , \ \ a_0^\pm = \frac{S^1 \mp i S^2}{\sqrt{2}} \nonumber\\
&& \eta=P^0\ , \ \  \pi^0 = P^3\ , \ \ \pi^\pm = \frac{P^1\mp iP^2}{\sqrt{2}}\ , \nonumber\\
&& B = \frac{B^5-iB^4}{\sqrt{2}}\ , \ \ \bar{B} =  \frac{B^5+iB^4}{\sqrt{2}}\ , \nonumber\\
&& B' = \frac{B'^5-iB'^4}{\sqrt{2}}\ , \ \ \bar{B}' =  \frac{B'^5+iB'^4}{\sqrt{2}} \ ,
\end{eqnarray}
and some generators of $U(4)$ as
\begin{eqnarray}
X^a &=& \frac{1}{2\sqrt{2}}\left(
\begin{array}{cc}
\tau_f^a & 0 \\
0 &(\tau_f^a)^T \\
\end{array}
\right)\ \ \ \ \ \ (a=0-3) \ , \nonumber\\
 X^i &=& \frac{1}{2\sqrt{2}}\left(
\begin{array}{cc}
0& D_f^i \\
(D_f^i)^\dagger & 0 \\
\end{array}
\right) \ \ \ \ \ \ (i=4,5) \ , \label{XMatrix}
\end{eqnarray}
with $\tau_f^0 = {\bm 1}$, $D_f^4=\tau_f^2$ and $D_f^5=i\tau_f^2$, the matrix~(\ref{SigmaPAssign}) turns into
\begin{eqnarray}
\Sigma =  (S^a- iP^a)X^aE + (B'^i - i B^i)X^iE\ . \label{SigmaRed}
\end{eqnarray}
The flavor matrix~(\ref{SigmaRed}) together with its transformation property~(\ref{SigmaTrans}) enables us to construct the linear sigma model in a familiar way but now based on the Pauli-Gursey $SU(4)$ symmetry of QC$_2$D.

\subsection{Linear sigma model}
\label{sec:LSM}

In this subsection we construct the linear sigma model from the flavor matrix $\Sigma$, which allows us to investigate the hadron mass spectrum at finite quark chemical potential.

From the flavor matrix~(\ref{SigmaRed}) with the transformation property~(\ref{SigmaTrans}), our linear sigma model that approximately preserves the Pauli-Gursey $SU(4)$ symmetry can be obtained as\footnote{Due to $\Sigma^\dagger = -\Sigma^*$, for instance, ${\rm tr}[\Sigma^*\Sigma]$ is identical to $-{\rm tr}[\Sigma^\dagger\Sigma]$.}
\begin{eqnarray}
{\cal L}_{\rm LSM} &=& {\rm tr}[D_\mu \Sigma^\dagger D^\mu\Sigma]-m_0^2{\rm tr}[\Sigma^\dagger\Sigma]-\lambda_1\big({\rm tr}[\Sigma^\dagger\Sigma]\big)^2 \nonumber\\
&-&\lambda_2{\rm tr}[(\Sigma^\dagger\Sigma)^2]+{\rm tr}[H^\dagger\Sigma+\Sigma^\dagger H] \nonumber\\
&+&c( {\rm det}\Sigma + {\rm det}\Sigma^\dagger) \ .
 \label{LSMTwoColor}
\end{eqnarray}
In Eq.~(\ref{LSMTwoColor}), we have left the flavor matrices up to the fourth order in $\Sigma$ ($\Sigma^\dagger$) such that the theory are renormalizable as widely done for the three-color version of linear sigma model~\cite{Lenaghan:2000ey,Schaefer:2008hk,Parganlija:2012fy}. $H$ is defined by
\begin{eqnarray}
H = h_q { E},
\end{eqnarray}
which describes the explicit breaking of the chiral symmetry or the Pauli-Gursey $SU(4)$ symmetry. 
Here, $h_q$ is a constant which captures the effects of the current quark masses.

Besides, the $U(1)_A$ axial transformation for $\Sigma$ is $\Sigma\to {\rm e}^{-i\theta_A{ I}}\Sigma{\rm e}^{-i\theta_A{ I}}$ as can be understood from Eq.~(\ref{AxialPsi}), and hence the Kobayashi-Maskawa-'t Hooft (KMT) type term proportional to $c$ is responsible for the $U(1)_A$ axial anomaly~\cite{Kobayashi:1970ji,Kobayashi:1971qz,tHooft:1976snw,tHooft:1976rip}. The covariant derivative with respect to $U(1)_B$ symmetry in Eq.~(\ref{LSMTwoColor}) is defined by
\begin{eqnarray}
D_\mu \Sigma = \partial_\mu\Sigma-i({\cal V}_\mu\Sigma+\Sigma {\cal V}_\mu^T) \ ,
\end{eqnarray}
where the ``gauge field'' ${\cal V}_\mu$ is replaced by
\begin{eqnarray}
{\cal V}_\mu = { J}\mu_q\delta_{\mu0}\ ,
\end{eqnarray}
with $\mu_q$ the quark number chemical potential introduced to access finite density.

In the vacuum the approximate Pauli-Gursey $SU(4)$ symmetry is further broken due to the VEV of chiral condensate $\langle\bar{\psi}\psi\rangle$, which is described by the appearance of a mean field of $\sigma$ in our model. In addition, at finite $\mu_q$ it is possible that the diquark condensate $\langle \psi^TC\gamma_5\tau_c^2\tau_f^2\psi\rangle$ emerges, leading to the baryon superfluidity that breaks the quark number conservation~\cite{Kogut:1999iv,Kogut:2000ek}. Such superfluidity is in our model triggered by a nonzero mean field of $B$ ($\bar{B}$). In fact, once, in Eq.~(\ref{LSMTwoColor}), one replaces $\sigma$ and $B^5$ by their mean fields which are real:\footnote{In this phase choice for $\Delta$, mean fields of $B$ and $\bar{B}$ become $\langle B\rangle = \langle\bar{B}\rangle = \Delta/\sqrt{2}$, and $B^4$ turns into the NG mode associated with the breakdown of baryon number symmetry.}
\begin{eqnarray}
\sigma_0 \equiv \langle\sigma\rangle\ , \ \  \Delta \equiv \langle B^5\rangle\ , \label{MeanFields}
\end{eqnarray}
the effective potential with respect to $\sigma_0$ and $\Delta$ can be obtained as
\begin{eqnarray}
V_{\sigma_0,\Delta} &=& -2\mu_q^2\Delta^2 + \frac{m_0^2}{2}(\sigma_0^2+\Delta^2) \nonumber\\
&+& \frac{8\lambda_1+2\lambda_2-c}{32}(\sigma_0^2+\Delta^2)^2 -2\sqrt{2}h_q\sigma_0\ . \label{PotentialV}
\end{eqnarray}
It should be noted that both the mean fields $\sigma_0$ and $\Delta$ keep the parity and isospin symmetries intact.

The mass of each hadron can be determined by expanding the Lagrangian~(\ref{LSMTwoColor}) up to quadratic order in the corresponding hadron field on top of the mean fields~(\ref{MeanFields}). We display their detailed expressions in Appendix~\ref{sec:MassDense} and here we only explain important features:
\begin{itemize}
\item In the vacuum where $\mu_q=0$ and naturally $\Delta=0$, a mass difference between $\pi$ and $\eta$ is proportional to $c$ that stems from the $U(1)_A$ axial anomaly as seen in Eq.~(\ref{MEtaVac}). In other words, in our model the mass of $\eta$ is pushed up by the anomaly effect as observed from the KMT term in three-color QCD~\cite{Kobayashi:1970ji,Kobayashi:1971qz,tHooft:1976snw,tHooft:1976rip}.

\item  For $\lambda_1=c=0$,\footnote{It corresponds to the leading approximation of the large $N_c$ expansion~\cite{Witten:1979kh,DiVecchia:1980yfw} as shown in Appendix~\ref{sec:NcCounting}.} the vacuum masses of $\eta$, $\pi$, $B$, and $\bar{B}$, which belong to the same multiplet of $SU(4)$, are degenerate, and so are those of $\sigma$, $a_0$, $B'$, and $\bar{B}'$ [see Eqs.~(\ref{VMass1}) and~(\ref{VMass2})]. These degeneracies indicate that effects of $SU(4)$ symmetry partly remain even when the symmetry is explicitly broken by current quark masses.

\item In the baryon superfluid phase where $\Delta$ is nonzero, $\sigma$, $B$, and $\bar{B}$, whose spin and parity are $J^P=0^+$, can mix. Similarly, $\eta$, $B'$, and $\bar{B}'$ having $J^P=0^-$ can mix in the superfluid phase. Such mixing stems from violation of baryon number conservation triggered by the diquark condensates. In fact, as can be seen from Eqs.~(\ref{MBSigma}) and~(\ref{MBpEta}) the corresponding mixing terms are proportional to $\Delta$. However, the mixing terms happen to be proportional to $\sigma_0$ as well, and therefore at sufficiently large $\mu_q$ when the chiral condensate becomes small due to the approximate restoration of chiral symmetry, all mixings are small too.
\end{itemize}

The ground state is determined by stationary conditions of the potential~(\ref{PotentialV}) with respect to $\sigma_0$ and $\Delta$. That is, the relevant mean fields must satisfy
\begin{eqnarray}
m_0^2+\frac{8\lambda_1+2\lambda_2-c}{8}(\sigma_0^2+\Delta^2) = \frac{2\sqrt{2}h_q}{\sigma_0}  \, , \label{GapEqSigma}
\end{eqnarray}
and
\begin{eqnarray}
\left(-4\mu_q^2+m_0^2+\frac{8\lambda_1+2\lambda_2-c}{8}(\sigma_0^2+\Delta^2)\right)\Delta = 0\ , \label{GapEqDelta}
\end{eqnarray}
respectively. Chiral symmetry or, more precisely, $SU(4)$ Pauli-Gursey symmetry is explicitly broken due to the current-quark mass effect $h_q$, and hence the trivial solution of $\sigma_0=0$ denoting the $SU(4)$ symmetric phase does not satisfy Eq.~(\ref{GapEqSigma}). On the other hand, Eq.~(\ref{GapEqDelta}) possesses both the trivial and nontrivial solutions of $\Delta$. The nontrivial solutions are selected by the value of chemical potential. In fact, once one inserts Eq.~(\ref{GapEqSigma}) into Eq.~(\ref{GapEqDelta}), the nontrivial $\Delta$ solution leads to
\begin{eqnarray}
 \mu_q^2 =  \frac{h_q}{\sqrt{2}\sigma_0}\ , \label{DeltaCodition}
 \end{eqnarray}
which cannot hold for smaller $\mu_q$. For adequately small $\mu_q$, therefore, the nontrivial solution of $\Delta$ can be discarded and hence the baryon superfluid phase does not emerge as naively expected. The trivial solution $\Delta=0$ leads to the hadronic phase which is continuously connected to the system with vanishing $\mu_q$. In the hadronic phase, according to Eq.~(\ref{GapEqSigma}), the value of $\sigma_0$ does not change from that in the vacuum $\sigma_0^{\rm vac}$. Note that the vacuum pion mass can be expressed as
\begin{eqnarray}
(m_\pi^{\rm vac})^2 = \frac{2\sqrt{2}h_q}{\sigma_0^{\rm vac}} \label{MPiVac}
\end{eqnarray}
from Eq.~(\ref{PiMassQC2D}).  Then, the critical chemical potential $\mu_q^*$ for the baryon superfluid phase can be analytically evaluated as
\begin{eqnarray}
\mu_q^* = \frac{m_\pi^{\rm vac}}{2}\ , \label{MuCritical}
\end{eqnarray}
from Eq.~(\ref{DeltaCodition}) with $\sigma_0$ being replaced by $\sigma_0^{\rm vac}$. The critical chemical potential~(\ref{MuCritical}) is the same as the result of chiral perturbation theory~\cite{Kogut:1999iv,Kogut:2000ek} and NJL model~\cite{Ratti:2004ra}\footnote{It is expected that the critical chemical potential~(\ref{MuCritical}) holds to all orders in perturbation theory. In particular, Eq.~(\ref{MuCritical}) was proven at one-loop order explicitly in chiral perturbation theory~\cite{Splittorff:2001fy}.}, and suggested numerically by lattice simulations~\cite{Hands:2001ee,Braguta:2016cpw,Iida:2019rah}.


The baryonic density can be evaluated by taking a derivative of the potential $V_{\sigma_0,\Delta}$ with respect to $\mu_q$:
\begin{eqnarray}
{\rho} = -\frac{\partial V_{\sigma_0,\Delta}}{\partial\mu_q} = 4\Delta^2\mu_q\ . 
\end{eqnarray}
Therefore, the baryonic density arises above the critical chemical potential $\mu_q$ accompanied by the onset of baryon superfluidity, whereas in the hadronic phase $\rho$ always vanishes. The latter constant behavior is related to the Silver Blaze property, which dictates the constancy of all thermodynamic quantities.

\begin{figure*}[t]
  \begin{center}
    \begin{tabular}{cc}

      \begin{minipage}[c]{0.5\hsize}
       \centering
       \hspace*{-2.5cm} 
         \includegraphics*[scale=0.6]{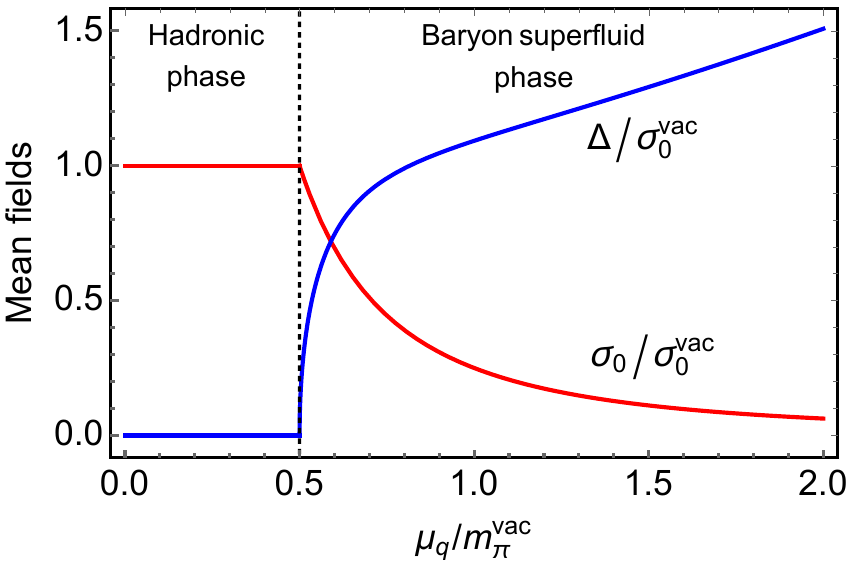}\\
         \end{minipage}

      \begin{minipage}[c]{0.4\hsize}
       \centering
        \hspace*{-1.1cm} 
          \includegraphics*[scale=0.49]{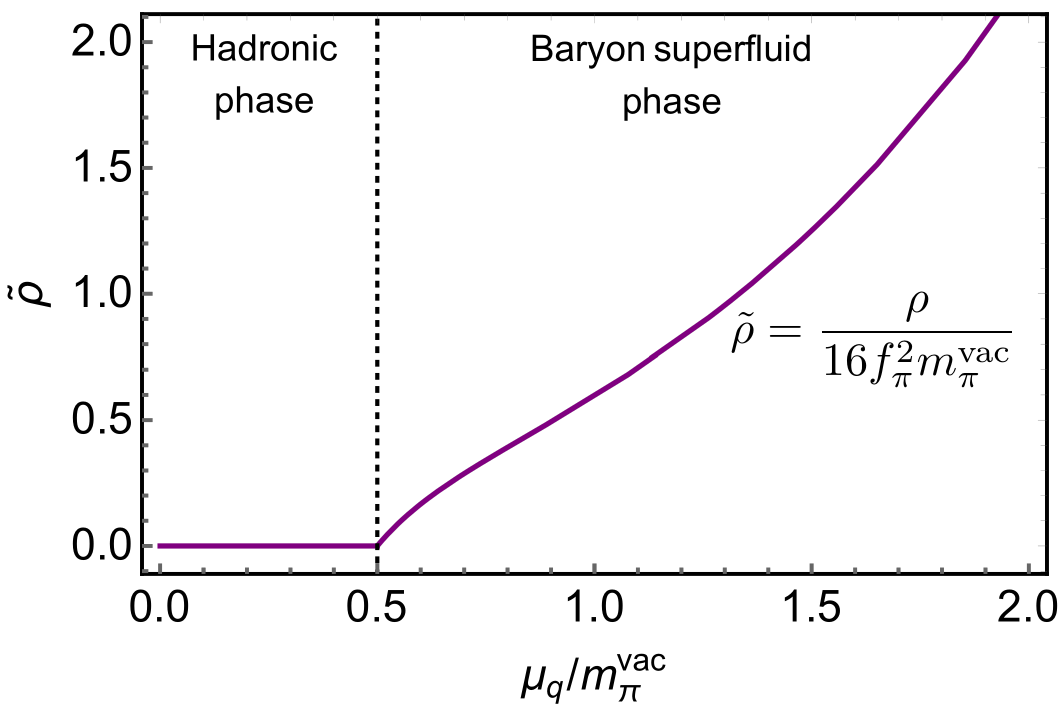}\\
      \end{minipage}

      \end{tabular}
 \caption{The $\mu_q$ dependence of $\sigma_0$ and $\Delta$ (left panel) and that of the scaled baryonic density $\tilde{\rho}$ (right panel). The vertical line corresponds to $\mu_q^*=m_\pi^{\rm vac}/2$, which distinguishes the normal and baryon superfluid phases.} 
\label{fig:MeanField}
  \end{center}
\end{figure*}

\section{Inputs}
\label{sec:Input}

In order to fix the model parameters, in the present work we employ the recent lattice results for hadron mass spectrum~\cite{Murakami:2022lmq,Murakami2022}, which ensures quantitatively convincing investigation.\footnote{In Ref.~\cite{Iida:2020emi}, the physical scale is fixed by $T_c=200$ MeV at $\mu_q=0$, where $T_c$ denotes the pseudo-critical temperature of the chiral phase transition.} Results from the lattice simulation, which are in part still tentative, are summarized as follows:
\begin{itemize}
\item In the physical unit, the pion mass is estimated to be $m_\pi=738$ MeV with good accuracy.
\item It seems that masses of $\pi$ and $\eta$ are almost identical in the hadronic phase, and hence we can take $c=0$. This choice implies disappearance of $U(1)_A$ anomaly effect in the hadronic phase.\footnote{In the simulation contributions from disconnected diagrams are not included. Such effects however seem to be negligible~\cite{Hands:2007uc,Murakami2022}.}
\item The measured masses of negative-parity baryons $B'$ and $\bar{B}'$ in the vacuum read $m_{B'(\bar{B}')}^{\rm vac}=[1611\pm128({\rm stat})_{-678}^{+128}({\rm syst})]\, {\rm MeV}$. Taking the central value as $m_{B'(\bar{B}')}^{\rm vac}\approx 1611\, {\rm MeV}$, we estimate a mass ratio of $B'$ ($\bar{B}'$) and $\pi$ to be $m_{B'(\bar{B}')}^{\rm vac}/m_\pi^{\rm vac} \approx2.18$.\footnote{The simulated values of $m_{B'}$ and $m_{\bar{B}'}$ at $\mu_q=119\, {\rm MeV}$ in the hadronic phase are $m_{B'}=[1238\pm87.6({\rm stat})^{+112}_{-87.6}({\rm syst})]\, {\rm MeV}$ and $m_{\bar{B}'}=[1704\pm65.5({\rm stat})^{+98.2}_{-71.0}({\rm syst})]\, {\rm MeV}$, respectively. When estimating $m_{B'(\bar{B}')}^{\rm vac}\approx 1611\, {\rm MeV}$, the mass formula that diquark baryons must satisfy in the hadronic phase as will be shown in Eq.~(\ref{DiquarkMassN}) reads $m_{B'}\approx 1373\, {\rm MeV}$ and $m_{\bar{B}'}\approx 1849\, {\rm MeV}$ at $\mu_q=119\, {\rm MeV}$. Deviations between these theoretical values and the simulated values are not large unreasonably. }
\item The computed mass of the $0^+$ scalar meson in the hadronic phase is also accompanied by uncertainties. The results in the vacuum ($\mu_q=0$) are quite noisy, but those at finite $\mu_q$ in the hadronic phase are rather worth using as inputs: $m_\sigma=[1453\pm 84.7({\rm stat})^{+103}_{-84.7}({\rm syst})]\, {\rm MeV}$ at $\mu_q=119\, {\rm MeV}$ and $m_\sigma=[1452\pm 101({\rm stat})^{+109}_{-106}({\rm syst})]\, {\rm MeV}$ at $\mu_q=238\, {\rm MeV}$. From these values and the assumption that $m_\sigma$ dose not change in the hadronic phase, we find $1.75\lesssim m_\sigma^{\rm vac}/m_\pi^{\rm vac}\lesssim 2.2$.
\end{itemize}
From those lattice inputs we can fix $m_0^2$, $c$, and $h_q$ and also determine a range of $\lambda_1$. In contrast, the dimensionless parameter $\lambda_2$ remains to be fixed. Then, we choose $\sigma_0^{\rm vac}=250$ MeV as a typical value to determine $\lambda_2$ in such a way that the magnitude of $\lambda_2$ becomes comparable to that broadly employed in the three-flavor linear sigma model~\cite{Lenaghan:2000ey,Schaefer:2008hk,Parganlija:2012fy}. The smaller (larger) value of $\sigma_0^{\rm vac}$ we take, the larger (smaller) value of $\lambda_2$ we obtain. We note that the choice of $\sigma_0^{\rm vac}$ does not affect the hadron mass spectrum at any $\mu_q$ as shown in Appendix~\ref{sec:NcCounting}, as long as we stick to the large $N_c$ limit, i.e., $\lambda_1=c=0$. The VEV $\sigma_0^{\rm vac}$ is related to the pion decay constant $f_\pi$ associated with the breakdown of $SU(4)\to Sp(4)$ as $f_\pi = \sigma_0^{\rm vac}/\sqrt{2}$; $f_\pi$ has yet to be measured on the lattice.

\begin{table}[htbp]
\begin{center}
  \begin{tabular}{l|ccccc} \hline
& $c$ & $\lambda_1$ & $\lambda_2$ & $m_0^2$ & $h_q$ \\ \hline 
Set (I) &  $0$ & $0$ & $65.6$  & $-(693\, {\rm MeV})^2$ & $(364\, {\rm MeV})^3$\\ 
Set (II) &  $0$ & $-7$ & $65.6$  & $-(206\, {\rm MeV})^2$ & $(364\, {\rm MeV})^3$\\ 
Set (III) &  $15$ & $0$ & $58.1$  & $-(495\, {\rm MeV})^2$ & $(364\, {\rm MeV})^3$\\
  \hline
 \end{tabular}
\caption{Parameter sets employed for computation of hadron mass spectrum in Sec.~\ref{sec:MassSpectrum}. The sets (I) and (II) are reasonable enough to reproduce the recent lattice results for hadron masses in the hadronic phase.}
\label{tab:Parameter}
\end{center}
\end{table}

From the above procedure, the range of $\lambda_1$ is found to be $-7\lesssim\lambda_1\lesssim0$, and hence for the numerical analysis in Sec.~\ref{sec:MassSpectrum} we consider two distinct cases $\lambda_1=0$ and $\lambda_1=-7$ for clear discussions. The resultant parameters are summarized in the sets (I) and (II) in Table~\ref{tab:Parameter}. In the table, although the simulated mass spectrum favors $c=0$, we also display the parameter set (III) with nonzero value of $c$ to examine effects of the $U(1)_A$ axial anomaly on hadron mass spectrum especially in the baryon superfluid phase later. Here, we again emphasize that the parameter set (I) where $\lambda_1=c=0$ is satisfied corresponds to the leading approximation of the large $N_c$ expansion.

\begin{figure*}[t]
  \begin{center}
    \begin{tabular}{cc}

      \begin{minipage}[c]{0.5\hsize}
       \centering
       \hspace*{-2.5cm} 
         \includegraphics*[scale=0.5]{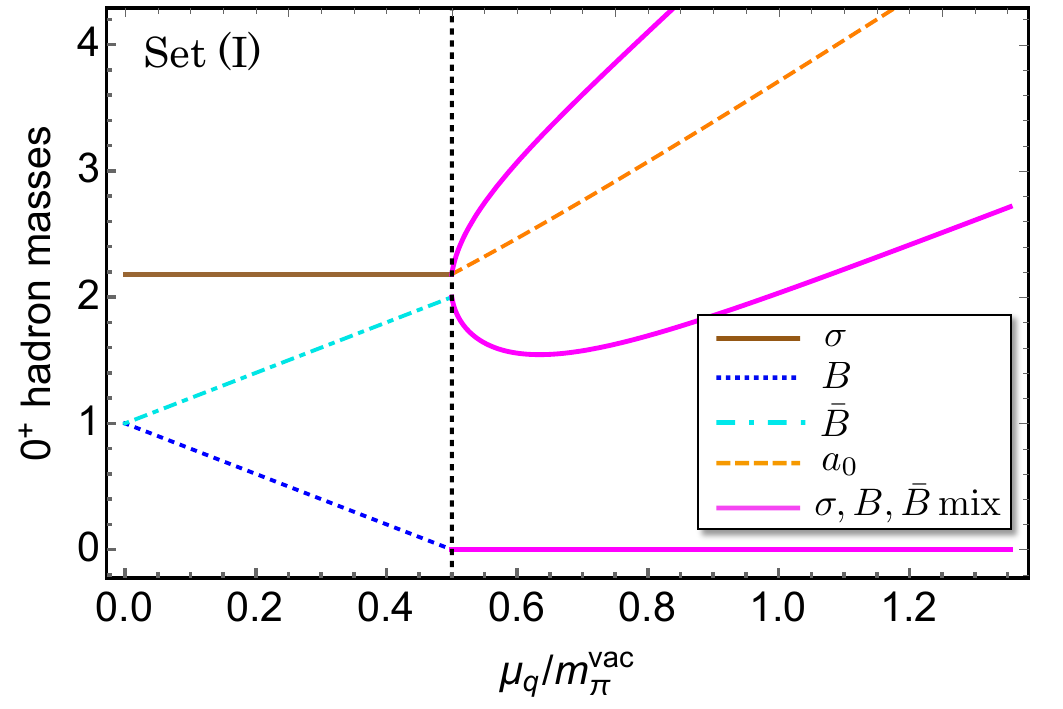}\\
         \end{minipage}

      \begin{minipage}[c]{0.4\hsize}
       \centering
        \hspace*{-1.1cm} 
          \includegraphics*[scale=0.5]{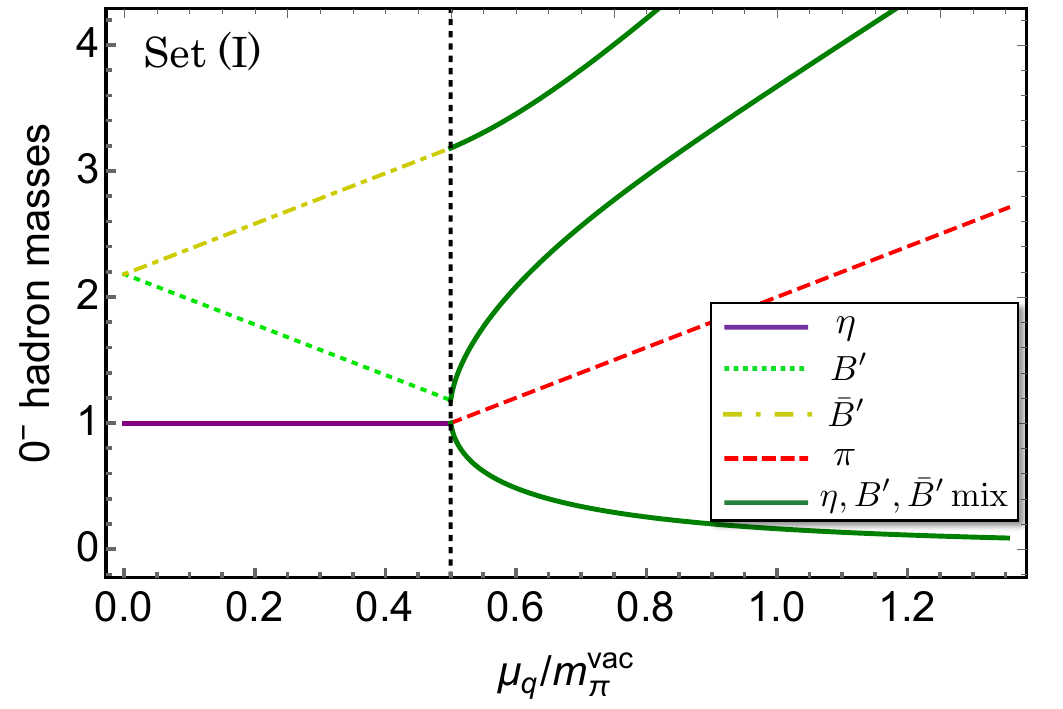}\\
      \end{minipage}

      \end{tabular}
 \caption{Left and right panels represent the $\mu_q$ dependence of $0^+$ hadron masses ($a_0$, $\sigma$, $B$, and $\bar{B}$) and that of $0^-$ hadron masses ($\pi$, $\eta$, $B'$, and $\bar{B}'$), respectively, calculated with the parameter set (I). These hadron masses are normalized by the vacuum pion mass $m_\pi^{\rm vac}$. The vertical dotted line corresponds to the critical chemical potential $\mu_q^*$.} 
\label{fig:Mass1}
  \end{center}
\end{figure*}

Before moving on to numerical computations of the hadron mass spectrum, we plot $\mu_q$ dependence of $\sigma_0$, $\Delta$, and $\rho$ in Fig.~\ref{fig:MeanField} with the parameter set (I) of Table~\ref{tab:Parameter} as a demonstration. The left panel depicts $\sigma_0$ (blue) and $\Delta$ (red) normalized by $\sigma_0^{\rm vac}$, respectively, and the right one depicts the scaled baryonic density~\cite{Hands:2001ee}
\begin{eqnarray}
\tilde{\rho} = \frac{\rho}{16f_\pi^2m_\pi^{\rm vac}}\ . \label{RhoTilde}
\end{eqnarray}
In the figure the vertical dotted line corresponds to the critical chemical potential $\mu_q^*$ given by Eq.~(\ref{MuCritical}), i.e., the transition between the hadronic and baryon superfluid phases. Figure~\ref{fig:MeanField} clearly exhibits the Silver Blaze property in the hadronic phase. Besides, the figure indicates that $\sigma_0$ decreases with $\mu_q$ in the baryon superfluid phase, resulting in the restoration of chiral symmetry at sufficiently high baryonic density. On the other hand, $\Delta$ increases monotonically as $\mu_q$ becomes large in the superfluid phase. Besides, unlike analysis from chiral perturbation theory within the mean field approach where $\sigma_0^2+\Delta^2=({\rm constant})$ is satisfied for any value of $\mu_q$~\cite{Kogut:1999iv,Kogut:2000ek},\footnote{The relation $\sigma_0^2+\Delta^2=({\rm constant})$ is violated when loop corrections are taken into account even in chiral perturbation theory.} the linear sigma model naturally violates such a conservation law. This is because the latter is based on the linear representation of quarks where the ground-state configuration is dynamically changed in accordance with the change of breaking strength of the Pauli-Gursey $SU(4)$ symmetry. A similar behavior is observed in the NJL model~\cite{Ratti:2004ra}.

\begin{figure*}[t]
  \begin{center}
    \begin{tabular}{cc}

      \begin{minipage}[c]{0.5\hsize}
       \centering
       \hspace*{-2.5cm} 
         \includegraphics*[scale=0.5]{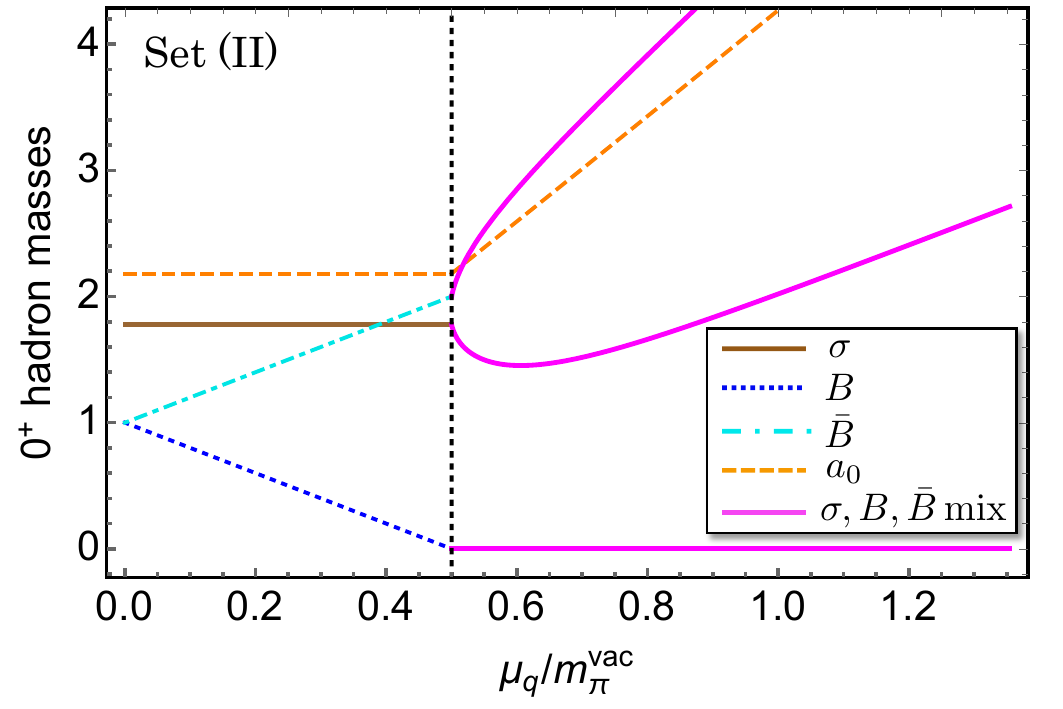}\\
         \end{minipage}

      \begin{minipage}[c]{0.4\hsize}
       \centering
        \hspace*{-1.1cm} 
          \includegraphics*[scale=0.5]{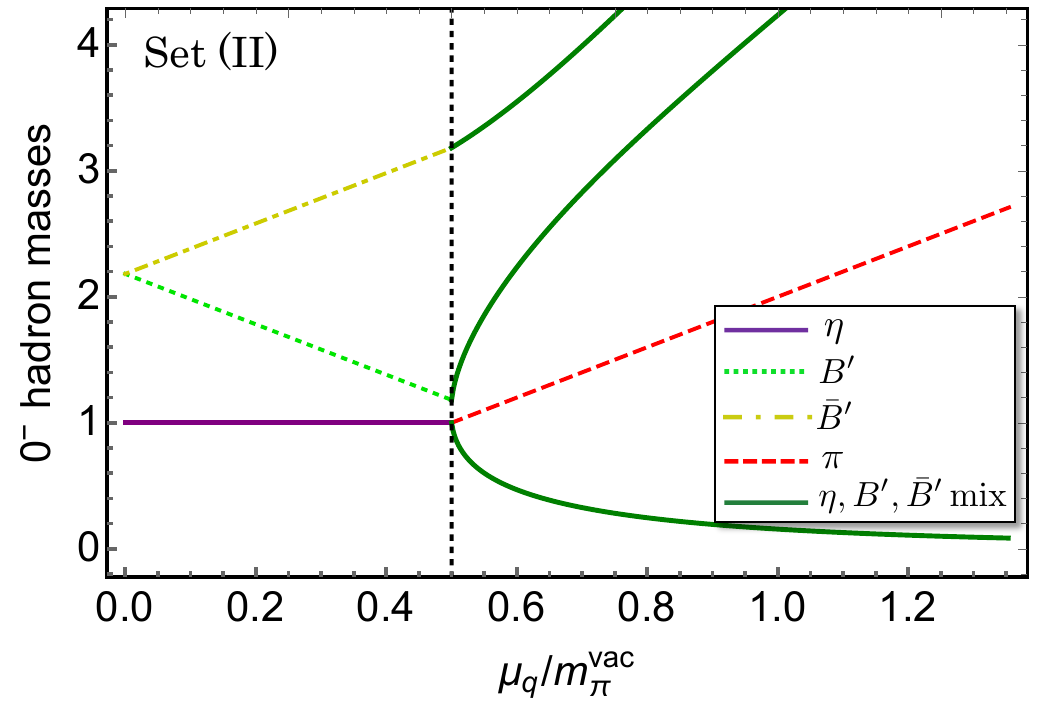}\\
      \end{minipage}

      \end{tabular}
 \caption{Same as Fig.~\ref{fig:Mass1} but with the parameter set (II).}
\label{fig:Mass2}
  \end{center}
\end{figure*}

\section{Mass spectrum}
\label{sec:MassSpectrum}

In this section we display the numerical results for $\mu_q$ dependence of the hadron masses evaluated in our present model. First, in Sec.~\ref{sec:SetI} we present the resultant mass spectrum with the parameter sets (I) and (II) in Table~\ref{tab:Parameter} consistent with the recent lattice simulation. Next, in Sec.~\ref{sec:SetIII} we study the mass modifications with the set (III) in Table~\ref{tab:Parameter} to have a closer look at effects of the $U(1)_A$ axial anomaly especially in the baryon superfluid phase. Finally in Sec.~\ref{sec:ChiralPartner}, we discuss the chiral partner structure in our model with the set (I).

\subsection{Results in the absence of $U(1)_A$ anomaly}
\label{sec:SetI}

Here, we investigate the hadron mass modifications for the parameter sets (I) and (II) in Table~\ref{tab:Parameter}, which are favored by the mass spectrum in the hadronic phase measured by the recent lattice simulation~\cite{Murakami:2022lmq,Murakami2022}. The sets are characterized by $c=0$, where the $U(1)_A$ anomaly is absent.

Depicted in Fig.~\ref{fig:Mass1} is the resultant $\mu_q$ dependence of the hadron masses with the set (I). The left and right panels indicate the $0^+$ ($a_0$, $\sigma$, $B$, and $\bar{B}$) and $0^-$ ($\pi$, $\eta$, $B'$, and $\bar{B}'$) hadron masses normalized by the vacuum pion mass $m_\pi^{\rm vac}$, respectively, and the vertical dotted line for these panels corresponds to the critical chemical potential $\mu_q^*$ in Eq.~(\ref{MuCritical}). As can be seen from the figure, in the hadronic phase for $\mu_q<\mu_q^*$ the masses of mesons, which do not carry the baryon number, are unchanged while those of baryons and antibaryons monotonically change as 
\begin{eqnarray}
m_{B,B'}&=&m_{B,B'}^{\rm vac}-2\mu_q \ , \nonumber\\
m_{\bar{B},\bar{B}'} &=& m_{\bar{B},\bar{B}'}^{\rm vac} + 2\mu_q\ . \label{DiquarkMassN}
\end{eqnarray}
These behaviors mean that in this phase the chemical potential $\mu_q$ simply shifts energy levels of the (anti)baryons without being accompanied by medium effects. Such stable $\mu_q$ dependences are understandable by the absence of baryonic density as in the right panel of Fig.~\ref{fig:MeanField}. 

Here, we note that for the parameter set (I), not only the $a_0$ and $\sigma$ masses but also the $\pi$ and $\eta$ ones are identical in the hadronic phase. The former is realized by the large $N_c$ condition $\lambda_1=c=0$, and the latter is solely by the neglect of $U(1)_A$ anomaly, i.e., $c=0$ as already explained. These properties are clearly understood by Fig.~\ref{fig:Mass2} where the mass spectrum is obtained with the parameter set (II); the figure shows that the negative $\lambda_1$ acts to lower the $\sigma$ mass, leading to breaking of the degeneracy of ($a_0$, $\sigma$), while it does not destroy the mass degeneracy of ($\pi$, $\eta$). As a result, a level crossing between $\sigma$ and $\bar{B}$ takes place below the critical chemical potential $\mu_q^*$. From the figure, it is also found that the positive $\lambda_1$ acts to increase the $\sigma$ mass to break the degeneracy of $a_0$ and $\sigma$.

\begin{figure*}[t]
  \begin{center}
    \begin{tabular}{cc}

      \begin{minipage}[c]{0.5\hsize}
       \centering
       \hspace*{-2.5cm} 
         \includegraphics*[scale=0.5]{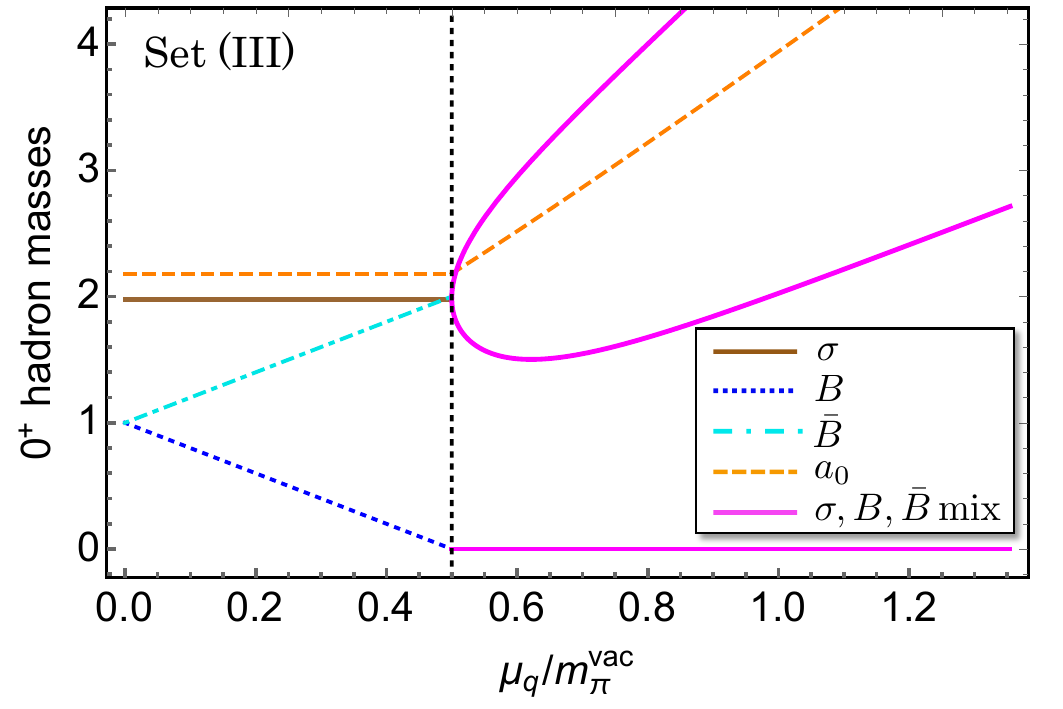}\\
         \end{minipage}

      \begin{minipage}[c]{0.4\hsize}
       \centering
        \hspace*{-1.1cm} 
          \includegraphics*[scale=0.5]{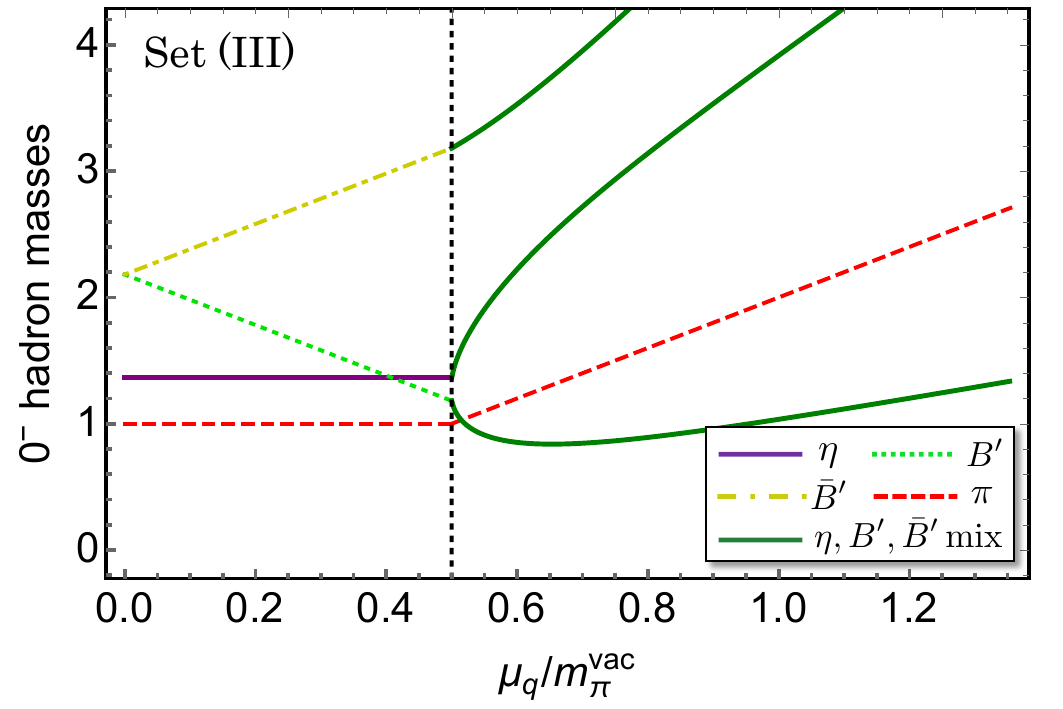}\\
      \end{minipage}

      \end{tabular}
 \caption{Same as Fig.~\ref{fig:Mass1} but with the parameter set (III).}
\label{fig:Mass3}
  \end{center}
\end{figure*}

In the baryon superfluid phase at $\mu_q>\mu_q^*$ in Figs.~\ref{fig:Mass1} and~\ref{fig:Mass2}, due to the violation of baryon number conservation several nontrivial mass modifications are found. First, for $J^P=0^+$ hadrons, $\sigma$, $B$, and $\bar{B}$ mix, and the lowest branch becomes massless, which plays the role of the NG boson associated with the violation of $U(1)_B$ baryon number symmetry. The isotriplet $a_0$ meson does not join the mixing since the $SU(2)$ isospin symmetry is not broken by $\Delta$. Next, for $J^P=0^-$ hadrons, $\eta$, $B'$, and $\bar{B}'$ also mix to draw a complicated mass spectrum. Due to the level repulsion among them, the lowest branch is pushed down and its mass becomes smaller than $m_\pi^{\rm vac}$.  Such remarkable behaviors of the lowest branches are certainly observed in the recent lattice simulation~\cite{Murakami:2022lmq,Murakami2022}. Comparing Fig.~\ref{fig:Mass1} and~\ref{fig:Mass2}, we can see that in the baryon superfluid phase the negative $\lambda_1$ acts to slightly increase the mass of $a_0$ and that of the second-lowest branch of the $\eta$-$B'$-$\bar{B}'$ mixed state. The mass orderings in the superfluid phase are nevertheless identical for the sets (I) and (II). For $\lambda_1=c=0$, it should be noted that the lowest branch of the $\eta$-$B'$-$\bar{B}'$ mixed state is reduced to a massless mode at sufficiently large $\mu_q$. In Sec.~\ref{sec:ChiralPartner} we will come back to this point.

 In addition to the above findings, interestingly enough, the $\mu_q$ dependence of the $\pi$ mass in the baryon superfluid phase is found to be expressed as 
\begin{eqnarray}
m_\pi^2 = 4\mu_q^2\ , \label{MPiSF}
\end{eqnarray}
which is the same as that predicted by chiral perturbation theory and the NJL model~\cite{Kogut:1999iv,Kogut:2000ek,Ratti:2004ra}. The mass formula~(\ref{MPiSF}) is analytically derived for any parameter set in the present model.

\subsection{Effects of the $U(1)_A$ anomaly}
\label{sec:SetIII}

As mentioned above, the mass spectrum in the hadronic phase measured by the recent lattice simulation supports the absence of $U(1)_A$ anomaly. Even so, it is useful to study the hadron mass spectrum at finite $\mu_q$ in the case in which the anomaly is present. For this reason, in this subsection we work with the parameter set (III).

Depicted in Fig.~\ref{fig:Mass3} is the result with the set (III). In the hadronic phase, the $U(1)_A$ anomaly effect with $c>0$ pushes down the $\sigma$ mass and pushes up the $\eta$ mass, resulting in breaking of mass degeneracies of $(a_0, \sigma)$ and of $(\pi, \eta)$. As a result, level crossings of $\bar{B}$ and $\sigma$ for $0^+$ hadrons and of $B'$ and $\eta$ for $0^-$ ones can occur below $\mu_q^*$. In the baryon superfluid phase, the mass spectrum is mostly similar to the one presented with the parameter sets (I) and (II) except the lowest branch of the $\eta$-$B'$-$\bar{B}'$ mixed state; notably the mass reduction of the state observed in Figs.~\ref{fig:Mass1} and~\ref{fig:Mass2} is tempered, and the mass again increases gradually well above $\mu_q^*$. Such a characteristic behavior is obviously distinct from the $c=0$ case where the mass reduction is striking and becomes asymptotically zero. Therefore, we conclude that such weakened mass reduction can be a useful signal to measure the change of the magnitude of the $U(1)_A$ anomaly in the superfluid phase.

\begin{figure}[t]
\centering
\hspace*{-0.5cm} 
\includegraphics*[scale=0.53]{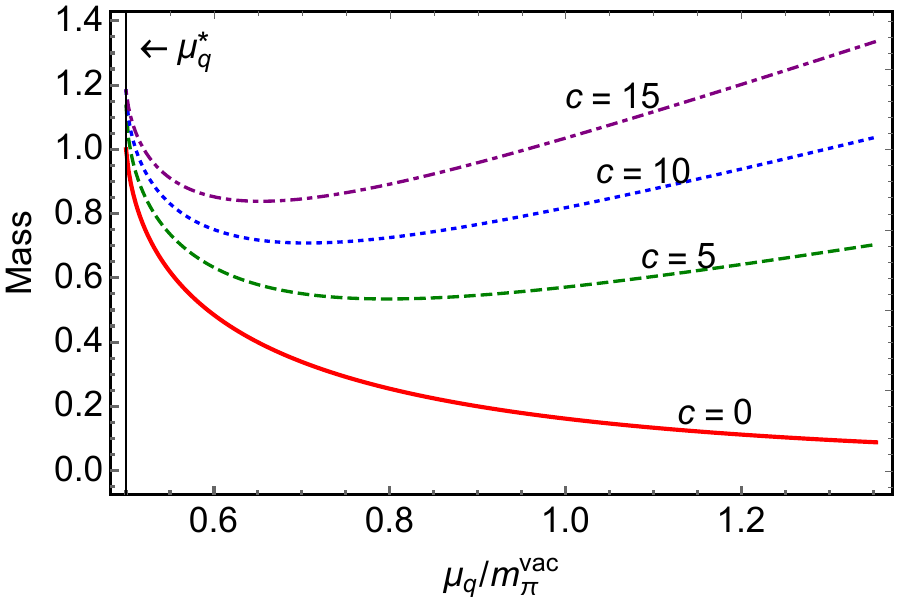}
\caption{The $\mu_q$ dependence of the mass of the lowest branch of the $\eta$-$B'$-$\bar{B}'$ mixed state in the baryon superfluid phase calculated for several $c$'s. The mass is normalized by $m_\pi^{\rm vac}$.}
\label{fig:Anomaly}
\end{figure}

In order to have a closer look at the influence of the $U(1)_A$ anomaly on the mass of the lowest branch of the $\eta$-$B'$-$\bar{B}'$ mixed state, we depict $\mu_q$ dependence of its mass in the baryon superfluid phase for several values of $c$ with $\lambda$ being set to zero in Fig.~\ref{fig:Anomaly}. From the figure one can see that the mass is strongly affected by the value of $c$, i.e., the magnitude of the $U(1)_A$ anomaly. In particular, when $c>0$ the mass is proportional to $\mu_q$ in the limit of $\mu_q\to\infty$, while only when $c=0$ the mass is reduced to zero in this limit.

\subsection{Chiral partner structures}
\label{sec:ChiralPartner}

One of the characteristic features of the conventional linear sigma model is manifestation of the so-called chiral partner structure~\cite{Hatsuda:1994pi}. That is, the linear representation of hadrons allows us to explore how the mass degeneracy occurs among hadrons having opposite parities via axial transformations at the chiral restoration point. In order to identify the mass degeneracy structure at large $\mu_q$, in this subsection we study the positive and negative parity hadrons simultaneously.

\begin{figure}[t]
\centering
\hspace*{-0.5cm} 
\includegraphics*[scale=0.62]{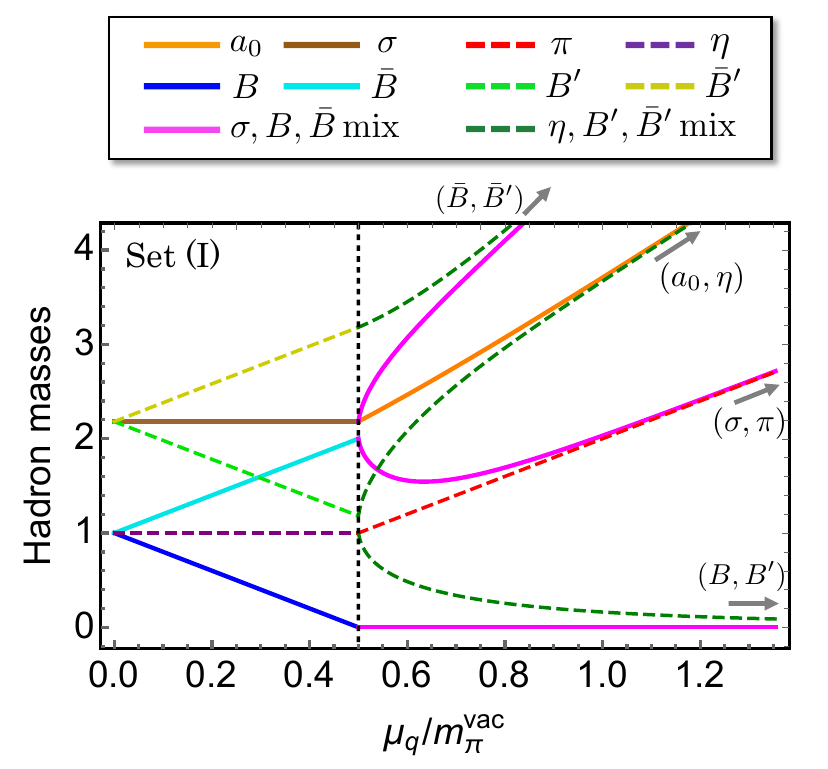}
\caption{The $\mu_q$ dependence of all the normalized hadron masses calculated with the parameter set (I). The solid and dashed curves denote the positive and negative  parity hadrons, respectively, and the vertical dotted line corresponds to $\mu_q^*$. The two states indicated in each parenthesis are the respective chiral partners realized at sufficiently large $\mu_q$.}
\label{fig:PlotAll}
\end{figure}

We display the $\mu_q$ dependence of all the hadron masses treated in the present model in Fig.~\ref{fig:PlotAll}. In this figure, we employ the parameter set (I) where the large $N_c$ limit is taken as the most instructive choice. The solid curves and dashed curves denote the positive and negative parity hadrons, respectively, and the vertical dotted line corresponds to $\mu_q^*$. Figure~\ref{fig:PlotAll} clearly shows that each state can be classified by asymptotic mass degeneracy as a chiral partner of the degenerate pair. Analytically, once one takes $\sigma_0\to0$ at large $\mu_q$ and sets $\lambda_1=c=0$ in the mass formulas in Appendix~\ref{sec:MassDense}, the troublesome mixing disappears and one can easily find
\begin{eqnarray}
&& m_{B}^2=m_{B'}^2 = 0 \ , \nonumber\\
&& m_\sigma^2 =m_\pi^2 = \mu_q^2 \ , \nonumber\\
&& m_{a_0}^2= m_\eta^2 = 12\mu_q^2 \  , \nonumber\\
&&  m_{\bar{B}}^2= m_{\bar{B}'}^2= 24\mu_q^2\ , \label{ChiralPartner}
\end{eqnarray}
which are independent of $\lambda_2$. The asymptotic mass formula~(\ref{ChiralPartner}) clearly exhibits the chiral partner structure. In particular, the formula indicates that the partners are $(B,B')$, $(\sigma,\pi)$, $(a_0,\eta)$, and $(\bar{B},\bar{B}')$ from the lowest mass, a sequence expected from the $SU(2)_A$ axial transformations. It should be noted that the chiral partner structures for baryons $(B,B')$ and antibaryons $(\bar{B},\bar{B}')$ are realized only when the $U(1)_A$ anomaly effect is switched off, i.e., $c=0$, and the large $N_c$ limit is taken, i.e., $\lambda_1=c=0$, respectively.

 The predicted mass degeneracies are expected to provide useful information of chiral symmetric properties of the hadrons at dense regime for future lattice simulations.

\section{Conclusions}
\label{sec:Conclusions}

In this article, motivated by the recent lattice simulation~\cite{Murakami2022,Murakami:2022lmq}, we have investigated hadron mass modifications at finite quark chemical potential $\mu_q$ in QC$_2$D with two flavors within the linear sigma model. The model enables us to study not only the masses of the ground-state pseudo-scalar mesons and scalar diquark baryons but also those of the chiral partners carrying opposite parities, which is the notable advantage of employing the linear sigma model. That is, we have succeeded in treating the positive parity mesons and diquark baryons: $\sigma$, $a_0$, $B$, and $\bar{B}$, and the negative parity ones: $\eta$, $\pi$, $B'$, and $\bar{B}'$, in a unified way.

In order to fix the model parameters, we have used the lattice results for hadron mass spectrum in the hadronic phase where the diquark condensate does not emerge~\cite{Murakami2022,Murakami:2022lmq}. In particular, the lattice result suggests that masses of $\pi$ and $\eta$ are identical in the hadronic phase, implying that effects of the $U(1)_A$ axial anomaly are suppressed there. Within our present model, such a suppression has been described by omitting the KMT type contributions.

In the baryon superfluid phase where the diquark condensate emerges, we have found a rich mass spectrum involving the mixing among $\sigma$-$B$-$\bar{B}$ for $J^P=0^+$ hadrons and that among $\eta$-$B'$-$\bar{B}'$ for $J^P=0^-$ ones, which is triggered by the $U(1)_B$ baryon number violation. The former mixing plays an essential role in describing the massless nature of the NG boson in association with the violation of $U(1)_B$ baryon number symmetry, while the latter leads to a noteworthy mass reduction of the lowest branch of the $\eta$-$B'$-$\bar{B}'$ mixed state. These characteristic properties have been indeed observed by the lattice simulation.

Besides, at sufficiently large $\mu_q$, we have demonstrated the chiral partner structure by deriving mass degeneracies of the hadrons that have opposite parities and are connected by the axial transformations. The predicted mass degeneracies are expected to be useful as guides for future lattice simulations toward elucidation of influence of the chiral restoration on the hadron properties, i.e., elucidation of the hadron mass generation.

In the absence of the $U(1)_A$ anomaly, the mass reduction of the lowest branch of the $\eta$-$B'$-$\bar{B}'$ mixed state is striking and the state finally becomes massless to exhibit the chiral partner structure with the NG boson. When the anomaly is present, the mass reduction is tempered such that the corresponding chiral partner structure is broken. The lattice simulation implies the latter tempered reduction. If this is the case, the $U(1)_A$ anomaly effects which are negligible in the hadronic phase possibly become sizable in the baryon superfluid phase. In this regard, we have also clarified a relation between the tempered mass reduction and the strength of the KMT determinant term. The relation is expected to be useful to derive the change of the magnitude of anomaly effects from further lattice simulations. Meanwhile, within effective models, the increment of the anomaly in medium measured by the magnitude of the KMT determinant term in three-color QCD was indeed reported in Refs.~\cite{Fejos:2016hbp,Fejos:2017kpq,Fejos:2018dyy}. Thus, we leave investigation on the strengthened anomaly effect in dense QC$_2$D matter for future study.

In what follows, we comment on relations between QC$_2$D and there-color QCD by focusing on the diquarks. In three-color QCD, diquarks themselves are not observable since they are not color singlets. Instead, singly heavy baryons consisting of one heavy quark and one diquark can be regarded as the corresponding hadrons to the diquark baryons in QC$_2$D. In three-color QCD, when one looks at the singly charmed baryons, the ground state is the well-established $\Lambda_c(2286)$~\cite{Workman:2022ynf} (counterpart of $B$ in QC$_2$D) but its chiral partner carrying a negative parity $\Lambda_c(\frac{1}{2}^-)$ (counterpart of $B'$ in QC$_2$D) has not yet observed experimentally despite theoretical predictions~\cite{Kawakami:2018olq,Kawakami:2019hpp,Harada:2019udr,Kim:2020imk,Kawakami:2020sxd,Suenaga:2021qri,Suenaga:2022ajn}. For this reason, seeking for $\Lambda_c(\frac{1}{2}^-)$ is one of the challenging topics on singly heavy baryon spectroscopy. On the other hand, in QC$_2$D the mass of negative-parity diquark $B$ ($\bar{B}$) has been certainly measured by the lattice simulation. However, the pion mass is rather heavy such that $B'$ ($\bar{B}'$) becomes stable. Therefore, numerical investigation of the hadrons in QC$_2$D by changing the pion mass, especially via dynamical aspects of $B'$ ($\bar{B}'$), would be desired to give clues to unveil problems on hidden $\Lambda_c(\frac{1}{2}^-)$ in three-color QCD. In addition, from further numerical elucidation of modifications of diquark baryons at finite density in QC$_2$D, it is expected that our deeper understanding of medium corrections of singly heavy baryons from a symmetry point of view would be achieved.

\section*{Acknowledgments}

The authors thank Makoto Oka for useful discussions. D.~S. is supported by the RIKEN special postdoctoral researcher program. K.~M. is supported in part by the Japan Society for the Promotion of Science (JSPS). E.~I. is supported by JSPS KAKENHI with Grant Number 19K03875, JST PRESTO Grant Number JPMJPR2113, JSPS Grant-in-Aid for Transformative Research Areas (A) JP21H05190 and JST Grant Number JPMJPF2221. K.~I. is supported by JSPS KAKENHI with Grant Number 18H05406.
The lattice Monte Carlo simulation is supported by the HPCI-JHPCN System Research Project (Project ID: jh220021).

\appendix

\section{Hadron masses at finite $\mu_q$}
\label{sec:MassDense}

Here, we show mass terms of the hadrons treated in our linear sigma model.

The mass terms are provided by expanding the Lagrangian~(\ref{LSMTwoColor}) up to quadratic order in the hadron fields, on top of the mean fields~(\ref{MeanFields}). 
The resultant Lagrangian reads
\begin{eqnarray}
{\cal L} &=& {\cal L}_{B}  +{\cal L}_\sigma  +{\cal L}_{B\sigma}  + {\cal L}_{B'} + {\cal L}_\eta+ {\cal L}_{B'\eta} \nonumber\\
&+& {\cal L}_{a_0} + {\cal L}_\pi + \cdots\ ,  \label{LSMTwoColor2}
\end{eqnarray}
where
\begin{eqnarray}
{\cal L}_B &=& \frac{1}{2}\partial_\mu B_4\partial^\mu B_4  +  \frac{1}{2}\partial_\mu B_5\partial^\mu B_5  + 2\mu_q(\partial_0B_4B_5\nonumber\\
&& - B_4\partial_0B_5) -\frac{m_{B_4}^2}{2}B_4^2 - \frac{m_{B_5}^2}{2}B_5^2\ , \\
{\cal L}_\sigma &=& \frac{1}{2}\partial_\mu \sigma\partial^\mu \sigma  - \frac{m_\sigma^2}{2}\sigma^2\ , \\  \nonumber\\ 
{\cal L}_{B\sigma} &=& -m_{B_5\sigma}^2 \sigma B_5\ , \label{SigmaBMix} \\  \nonumber\\ 
 {\cal L}_{B'} &=& \frac{1}{2}\partial_\mu B'_4\partial^\mu B'_4  +  \frac{1}{2}\partial_\mu B'_5\partial^\mu B'_5 + 2\mu_q(\partial_0 B'_4B'_5 \nonumber\\ 
 && - B'_4\partial_0 B'_5)  -\frac{m_{B'_4}^2}{2}B_4'^2 - \frac{m_{B'_5}^2}{2}B_5'^2\ , \\  \nonumber\\ 
 {\cal L}_\eta &=&  \frac{1}{2}\partial_\mu \eta\partial^\mu\eta -\frac{m_\eta^2}{2}\eta^2\ , \\  \nonumber\\ 
{\cal L}_{B'\eta} &=& -m_{B_5'\eta}^2 B_5'\eta\ , \label{BpEtaMix} \\  \nonumber\\ 
 {\cal L}_{a_0} &=&  \frac{1}{2}\partial_\mu a_0^a\partial^\mu a_0^a -\frac{m_{a_0}^2}{2}a_0^aa_0^a \ \ \ (a=1,2,3)\ , \\  \nonumber\\
 {\cal L}_\pi &=&  \frac{1}{2}\partial_\mu \pi^a\partial^\mu\pi^a -\frac{m_{\pi}^2}{2}\pi^a\pi^a\ \ \ (a=1,2,3)\ ,
 \end{eqnarray}
with the corresponding masses
\begin{eqnarray}
m_\pi^2 &=&  m_0^2 +\frac{8\lambda_1+2\lambda_2-c}{8}(\sigma_0^2+\Delta^2) \ ,
\label{PiMassQC2D} \\ \nonumber\\
m_{a_0}^2 &=&  m_\pi^2 + \frac{\lambda_2}{2}(\sigma_0^2+\Delta^2) + \frac{c}{4}(\sigma_0^2+\Delta^2)\ ,  \\ \nonumber\\
m_{B_4}^2 &=&  m_\pi^2 - 4\mu_q^2\ , \\ \nonumber\\
m_{B_5}^2 &=& m_\pi^2-4\mu_q^2 + \frac{8\lambda_1+2\lambda_2-c}{4}\Delta^2\ , \\ \nonumber\\
m_\sigma^2 &=& m_\pi^2 +\frac{8\lambda_1+2\lambda_2-c}{4}\sigma_0^2\ , \\  \nonumber\\
m_{B_5\sigma}^2 &=& \frac{8\lambda_1+2\lambda_2-c}{4}\sigma_0\Delta\ , \label{MBSigma} \\ \nonumber\\
m_{B_4'}^2 &=& m_\pi^2-4\mu_q^2 + \frac{2\lambda_2+c}{4}(\sigma_0^2+\Delta^2)\ , \\ \nonumber\\
m_{B_5'}^2 &=& m_\pi^2-4\mu_q^2 + \frac{\lambda_2}{2}\sigma_0^2+\frac{c}{4}(\sigma_0^2+2\Delta^2)  \ , \\ \nonumber\\
m_\eta^2 &=& m_\pi^2 + \frac{\lambda_2}{2}\Delta^2 + \frac{c}{4}(2\sigma_0^2+\Delta^2)\ , \label{MEta} \\ \nonumber\\
m_{B_5'\eta}^2 &=& \frac{2\lambda_2-c}{4}\sigma_0\Delta\ . \label{MBpEta}
\end{eqnarray} 
Equations~(\ref{SigmaBMix}) and~(\ref{BpEtaMix}) tell us that not only $\sigma$, $B$, and $\bar{B}$ but also $\eta$, $B'$, and $\bar{B}'$ can mix in the baryon superfluid phase where $\Delta\neq0$, due to the violation of baryon number conservation. In fact, as can be seen from Eqs.~(\ref{MBSigma}) and~(\ref{MBpEta}) those mixing terms are proportional to $\Delta$. 

In order to examine the detailed structure of hadron masses in our model, we focus on the vacuum described by $\mu_q=0$ with $\sigma_0=\sigma_0^{\rm vac}$ and $\Delta=0$. The resultant hadron masses read
\begin{eqnarray}
 (m_\pi^{\rm vac})^2 = (m_{B_4}^{\rm vac})^2 = (m_{B_5}^{\rm vac})^2\ ,
 \end{eqnarray}
\begin{eqnarray} 
 (m_\eta^{\rm vac})^2 = (m_\pi^{\rm vac})^2 + \frac{c}{2}(\sigma_0^{\rm vac})^2 \ , \label{MEtaVac}
 \end{eqnarray}
\begin{eqnarray}
&& (m_{a_0}^{\rm vac})^2 = (m_{B'_4}^{\rm vac})^2 = (m_{B'_5}^{\rm vac})^2 \nonumber\\
 &=&  (m_\pi^{\rm vac})^2+ \frac{2\lambda_2+c}{4}(\sigma_0^{\rm vac})^2\ ,  \label{MA0Vac}
\end{eqnarray}
\begin{eqnarray}
(m_\sigma^{\rm vac})^2 =  (m_\pi^{\rm vac})^2+ \frac{8\lambda_1+2\lambda_2-c}{4}(\sigma_0^{\rm vac})^2\ .
\end{eqnarray}
Equation~(\ref{MEtaVac}) implies that in the vacuum the mass difference between $\eta$ and $\pi$ is proportional to $c$ and thus stems from the $U(1)_A$ anomaly. Typically the $\eta$ meson is heavier than $\pi$, so in this case $c>0$. Besides, when assuming the large $N_c$ limit, i.e., $\lambda_1=c=0$, we can find
\begin{eqnarray}
(m_\eta^{\rm vac})^2 = (m_\pi^{\rm vac})^2 = (m_{B_4}^{\rm vac})^2 = (m_{B_5}^{\rm vac})^2\ , \label{VMass1}
\end{eqnarray}
and
\begin{eqnarray}
(m_{\sigma}^{\rm vac})^2 = (m_{a_0}^{\rm vac})^2 = (m_{B'_4}^{\rm vac})^2 = (m_{B'_5}^{\rm vac})^2 \ . \label{VMass2}
\end{eqnarray}

\section{The $N_c$ counting}
\label{sec:NcCounting}

In this Appendix, we give explanations of our $N_c$ counting of the model parameters and also clarify how hadron masses 
depend on the value of $\sigma_0^{\rm vac}$.

As is well known, diagrams in the mesonic level are of ${\cal O}(N_c)$ since the leading contributions are scaled in the same way as a simple quark loop when the gauge coupling $g_c$ is scaled as $N_c^{-1/2}$. Meanwhile, wave functions of the mesons are of ${\cal O}(\sqrt{N_c})$~\cite{Witten:1979kh,DiVecchia:1980yfw}. Thus, the $N_c$ counting of coupling constants in effective models involving $n$ mesons is estimated to be of ${\cal O}(N_c^{(2-n)/2})$~\cite{Parganlija:2012fy}. Within this $N_c$ counting, $m_0^2$ and $\lambda_2$ behave as $m_0^2 = {\cal O}(N_c^0)$ and $\lambda_2= {\cal O}(N_c^{-1})$, respectively. The other four-point coupling $\lambda_1$ is, however, scaled as $N_c^{-2}$ since the $\lambda_1$ term includes two traces with respect to flavors; the leading contributions cannot be described by one quark loop but by two loops mediated by gluons in between. Phenomenologically, such an $N_c$ suppression is referred to as the Zweig rule. Besides, the constant $h_q$, which is responsible for the explicit breaking of the Pauli-Gursey $SU(4)$ symmetry, is $h_q={\cal O}(N_c^{1/2})$, and $\sigma_0$ is of ${\cal O}(N_c^{1/2})$. By combining these $N_c$ countings with Eqs.~(\ref{MPiVac}),~(\ref{DeltaCodition}) and~(\ref{MPiSF}), the pion mass in both the vacuum and medium can be understood to be of ${\cal O}(N_c^0)$ as expected.

The $N_c$ counting of the anomalous contribution $c$ can be determined by focusing on the $\eta$ mass formula in the vacuum. As discussed in Ref.~\cite{Witten:1979kh}, the $\eta$ mass must be scaled as $N_c^{-1/2}$ in such a way that the $\eta$ meson turns into an NG boson in association with the suppression of the $U(1)_A$ anomaly. Therefore, we can conclude from Eq.~(\ref{MEtaVac}) that $c$ is scaled as $N_c^{-2}$. 

To summarize, our $N_c$ counting of the model parameters is determined as
\begin{eqnarray}
&& m_0^2={\cal O}(N_c^0)\ , \ \ \lambda_1={\cal O}(N_c^{-2})\ , \ \ \lambda_2 = {\cal O}(N_c^{-1})\ , \nonumber\\
&& h_q={\cal O}(N_c^{1/2})\ , \ \ c={\cal O}(N_c^{-2})\ .
\end{eqnarray}
Therefore, the parameter set with $\lambda_1=c=0$ corresponds to the large $N_c$ limit in which higher-order contributions can be discarded.

In the large $N_c$ limit where $\lambda_1=c=0$, one notable universal behavior of the hadron mass spectrum at finite $\mu_q$ can be derived. In this limit, from the stationary conditions for $\sigma_0$ and $\Delta$ in Eqs.~(\ref{GapEqSigma}) and~(\ref{GapEqDelta}), the nontrivial solutions are found to satisfy
\begin{eqnarray}
\lambda_2\sigma_0^2 &=& \left(\frac{m_\pi^{\rm vac}}{m_\pi}\right)^4\lambda_2(\sigma_0^{\rm vac})^2\ , \nonumber\\
\lambda_2\Delta^2 &=& \left(1-\frac{(m_\pi^{\rm vac})^2}{m_\pi^2}\right)\left[\lambda_2(\sigma_0^{\rm vac})^2+4m_\pi^2\right]\ , \label{SigmaDelta}
\end{eqnarray}
respectively. That is, when we take $m_\pi^{\rm vac}$ and $m_{B'(\bar{B}')}^{\rm vac}$ as inputs, 
\begin{eqnarray}
\lambda_2(\sigma_0^{\rm vac})^2 = 2(m_{B'(\bar{B}')}^{\rm vac})^2-2(m_\pi^{\rm vac})^2=({\rm constant})
\end{eqnarray}
holds from Eq.~(\ref{MA0Vac}), and accordingly $\lambda_2\sigma_0^2$ and $\lambda_2\Delta^2$ depend only on $\mu_q$ as can be seen from Eq.~(\ref{SigmaDelta}) with $m_\pi^2=4\mu_q^2$. On the other hand, the mass formulas~(\ref{PiMassQC2D})--(\ref{MBpEta}) become dependent only on $\lambda_2\sigma_0^2$ and $\lambda_2\Delta^2$ for $\lambda_1=c=0$. Therefore, the hadron masses in the baryon superfluid phase turn out to be dependent on $\mu_q$ alone in the large $N_c$ limit and hence unaffected by the vacuum value $\sigma_0^{\rm vac}$. In other words, the hadron mass spectrum in both the hadronic and baryon superfluid phases is independent of the choice of $\sigma_0^{\rm vac}$ in the limit of interest here.

\bibliography{reference}

\end{document}